\pdfoutput=1
\documentclass[10pt, conference, compsocconf]{IEEEtran}
\usepackage{siunitx}
\usepackage{color}

\newtheorem{theorem}{Theorem}

\usepackage{cite}

\ifCLASSINFOpdf
  \usepackage[pdftex]{graphicx}
\else
\fi
\usepackage[cmex10]{amsmath}
\usepackage{url}

\begin{document}
\onecolumn
© 2021 IEEE. Personal use of this material is permitted. Permission from IEEE must be obtained for all other uses, in any current or future media, including reprinting/republishing this material for advertising or promotional purposes, creating new collective works, for resale or redistribution to servers or lists, or reuse of any copyrighted component of this work in other works.\\

\url{DOI: 10.1109/ICIEA52957.2021.9436749}
\cleardoublepage
\twocolumn
%
\title{Larger Sparse Quadratic Assignment Problem Optimization\\ Using Quantum Annealing and a Bit-Flip Heuristic Algorithm}




%
\author{\authorblockN{Michiya Kuramata}
\authorblockA{Department of Industrial \\
Engineering and Economics\\
Tokyo Institute of Technology\\
Tokyo, Japan\\
Email: kuramata.m.aa@m.titech.ac.jp}
\and
\authorblockN{Ryota Katsuki}
\authorblockA{NTT DATA Corporation\\
Tokyo, Japan\\
Email: Ryota.Katsuki@nttdata.com}
\and
\authorblockN{Kazuhide Nakata}
\authorblockA{Department of Industrial \\
Engineering and Economics\\
Tokyo Institute of Technology\\
Tokyo, Japan\\
Email: nakata.k.ac@m.titech.ac.jp}}


\maketitle

\begin{abstract}
Quantum annealing and D-Wave quantum annealer attracted considerable attention for their ability to solve combinatorial optimization problems. In order to solve other type of optimization problems, it is necessary to apply certain kinds of mathematical transformations. However, linear constraints reduce the size of problems that can be represented in quantum annealers, owing to the sparseness of connections between qubits. For example, the quadratic assignment problem (QAP) with linear equality constraints can be solved only up to size 12 in the quantum annealer D-Wave Advantage, which has 5640 qubits. To overcome this obstacle, Ohzeki developed a method for relaxing the linear equality constraints and numerically verified the effectiveness of this method with some target problems, but others remain unsolvable. In particular, it is difficult to obtain feasible solutions to problems with hard constraints, such as the QAP. We therefore propose a method for solving the QAP with quantum annealing by applying a post-processing bit-flip heuristic algorithm to solutions obtained by the Ohzeki method. In a numerical experiment, we solved a sparse QAP by the proposed method. This sparse QAP has been used in areas such as item listing on an E-commerce website. We successfully solved a QAP of size 19 with high accuracy for the first time using D-Wave Advantage. We also confirmed that the bit-flip heuristic algorithm moves infeasible solutions to nearby feasible solutions in terms of Hamming distance with good computational efficiency.

\end{abstract}


%
\IEEEpeerreviewmaketitle

\section{Introduction}
\label{Introduction}
Quantum computers are expected to accelerate computation in fields such as combinatorial optimization, simulation, machine learning, and so on. There are two main types of quantum computers: gate-model quantum computers and quantum annealers. In this study, we focus on quantum annealers that are hardware implementations of quantum annealing \cite{kadowaki1998quantum}. Quantum annealers have started to gain practical computational capabilities, particularly in combinatorial optimization and sampling. D-Wave Systems has been developing quantum annealers with steadily increasing numbers of qubits. In 2011, D-Wave Systems released the first commercial quantum annealer, D-Wave One, which had 128 qubits. Since then, they released D-Wave Two with 512 qubits in 2013 and D-Wave 2X with 1152 qubits in 2015, followed by the quantum annealer D-Wave 2000Q with 2048 qubits in 2017 and D-Wave Advantage \cite{dwave2019advantage} with 5640 qubits in September 2020.

With this steady development of hardware, there has been an increasing interest in finding use cases. Some practical applications to real-world problems exist \cite{neukart2017traffic}\cite{nishimura2019item}, particularly in the field of combinatorial optimization. Neukart et al. \cite{neukart2017traffic} proposed an optimization problem for assigning cars to roads to avoid congestion, which was solved using D-Wave 2X. Nishimura et al. \cite{nishimura2019item} used D-Wave 2000Q to optimize the order in which hotels are listed on a Japanese hotel booking website (Jalan) as a quadratic assignment problem (QAP) \cite{koopmans1957assignment}. Both studies computed only small problems using quantum annealers and larger problems using Qbsolv \cite{dwave2017qbsolv}, an algorithm based on tabu search.

The current D-Wave quantum annealer has many limitations. One is that there are sparse connections between qubits, meaning there is no link between two distant qubits, so multiple qubits may be required to represent a single logical variable. Particularly in the case of a combinatorial optimization problem such as QAP, which has a one-hot constraint such that only one of multiple 0–1 binary variables takes a value of 1, a connection is required between any pair of logical variables. The quantum annealer D-Wave Advantage with 5640 qubits can thus only solve up to 180 fully connected logical variables if missing qubits are not considered. Therefore, the maximum QAP size that can be directly solved is 12 or less.

To overcome such obstacles, Ohzeki \cite{masayuki2020breaking} proposed a method that relaxes linear equality constraints such as one-hot constraints. Using this method, we can represent larger problems on the quantum annealer hardware. Ohzeki numerically verified the effectiveness of this method with some problems, but the target problems cannot always be solved. In particular, it is difficult to obtain feasible solutions of problems, such as QAP, as shown in Section \ref{RESULT AND DISCUSSION}.

\section{PREVIOUS STUDIES}
\label{PREVIOUS STUDIES}
\subsection{Quantum Annealing Using D-Wave Hardware }
\label{PREVIOUS STUDIES-Quantum Annealing Using D-Wave2000}
Like quantum annealing, simulated annealing (SA) \cite{kirkpatrick1983optimization} is a method for solving combinatorial optimization problems. SA incorporates a pseudo-concept of heat, which represents the ease of state change on a computer. Heat is initially high to promote the active state transition. Then, the solution with the minimum objective function value is found by gradually decreasing the heat. In contrast, quantum annealing \cite{kadowaki1998quantum} obtains the minimum value of an objective function by introducing and gradually weakening a transverse magnetic field instead of SA heat. Quantum annealing is not performed pseudo-operatively on a classical computer, but as a physical phenomenon in a quantum annealer, which performs quantum annealing according to a physical model called the transverse field Ising model, represented by
\begin{equation}
\label{eq:tranverse ising model}
\begin{split}
\mathcal{H}(t) &=\left(1-s(t)\right) \left(-\sum_{i,j} J_{i j} \sigma_{i}^{z} \sigma_{j}^{z}-h \sum_{i} \sigma_{i}^{z} \right) \\& + s(t) \left( -\sum_{i} \sigma_{i}^{x} \right),
\end{split}
\end{equation}
where $J_{ij}$ represents interaction of the $i$-th and $j$-th qubits, $\sigma_{i}^{z}$ is the Pauli Z operator in the $i$-th qubit, $\sigma_{i}^{x}$ represents the Pauli X operator in the $i$-th qubit, and $s(t)$ is a function of time $t$ that monotonically decreases from $1$ to $0$. In quantum annealing, decreasing $s(t)$ corresponds to gradual weakening of the transverse magnetic field, which finally gives the state $\lbrace s_{i} | i \in V \rbrace$ with a minimum value as an energy function of a physical model called the Ising model, namely,
\begin{equation}
\label{eq: ising model}
E=\sum_{i,j} J_{i j} s_{i} s_{j}+\sum_{i} h_{i} s_{i}, \quad s_i=\pm 1\;\; , \forall i \in V,
\end{equation}
where $J_{ij}$ represents the coupling parameter of the $i$-th and $j$-th qubits, $V$ is a set of qubits, and $h_i$ represents bias applied to the $i$-th qubit. The combinatorial optimization problem can be solved using physical hardware by setting $J_{ij}$ and $h_i$ so that the objective function of the combinatorial optimization problem and the energy function of the Ising model are equivalent. Handling combinatorial optimization problems in the form of an Ising model is complicated because they are formulated in terms of binary variables. Therefore, we represent the combinatorial optimization problem as the quadratic unconstrained binary optimization (QUBO) problem
\begin{equation}
\label{eq: QUBO}
\begin{array}{ll}
\text{minimize} & \mathbf{q}^{\top} Q \mathbf{q} \\
\text {subject to} & \mathbf{q} \in \lbrace0,1\rbrace^{L},
\end{array} 
\end{equation}
where $q_i$ is a binary variable, $L$ is the number of $q_i$, and $Q$ is an $L \times L$ matrix. QUBO is the problem of finding $\mathbf{q}$ that minimizes $\mathbf{q}^{\top}Q\mathbf{q}$, which is equivalent to the Ising model. In other words, a combinatorial optimization problem formulated as QUBO can be optimized using quantum annealing. We can obtain QUBO, substituting $s_i=2q_i-1$ into the energy function of the Ising model of (\ref{eq: ising model}). QUBO does not allow for constraints. Therefore, to ensure that a solution satisfies its constraints, QUBO needs a penalty term in the objective function. In this paper, the combinatorial optimization problem is also represented as QUBO and optimized by quantum annealing.

\begin{figure}[!t]
\centering
\includegraphics[width=3.3in]{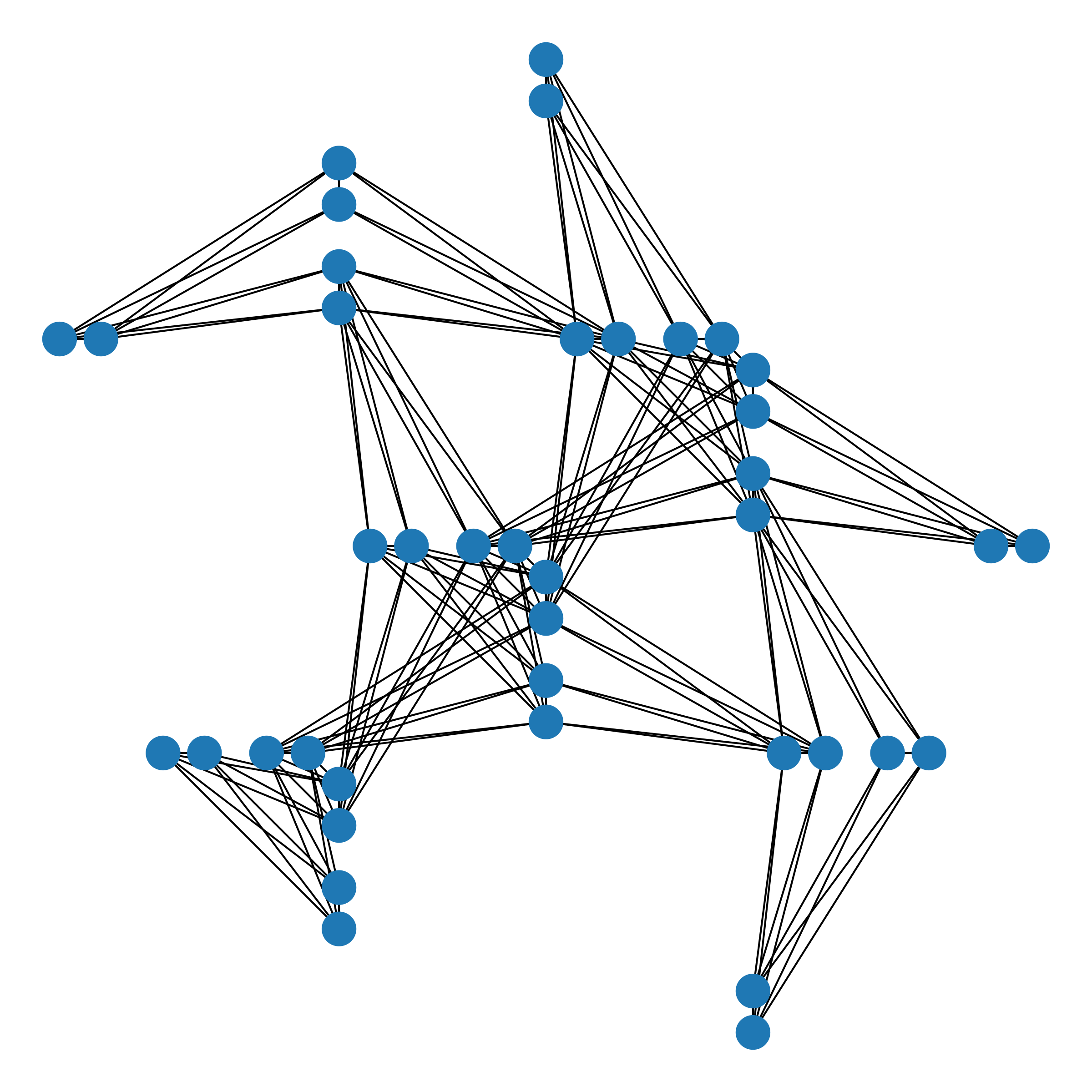}
\caption{Pegasus graph. Blue circles represent qubits, and lines connecting qubits represent connections between them.}             
\label{fig:pegasus}
\end{figure}

D-Wave Systems’ quantum processing unit (QPU) has a graph structure with qubits as nodes and connections between qubits as edges. D-Wave Advantage uses the Pegasus graph shown in Figure \ref{fig:pegasus}, where blue circles represent qubits and lines represent connections between qubits. One qubit does not have a connection with all other qubits, and connections between qubits are sparse. Therefore, representing a single logical variable may require multiple qubits. In other words, if we map multiple well-connected logical variables to the Pegasus graph, we find that they consume more qubits to represent a single logical variable. The number of connections between logical variables corresponds to the number of nonzero elements of $Q$ in (\ref{eq: QUBO}). Therefore, QUBO with $Q$ with many nonzero elements is harder to represent in the Pegasus graph, while QUBO with $Q$ with few nonzero elements is easier to represent. Particularly in the case of a combinatorial optimization problem with one-hot constraints, connections are needed between any pair of logical variables in those constraints. Therefore, D-Wave Advantage cannot solve problems of size 13 or larger, such as the QAP or the traveling salesman problem (TSP).

Note that in this paper, we define a one-hot constraint as one where exactly one of multiple binary variables takes value 1. In a QAP, for example, the constraint that a facility can be built at exactly one location is a one-hot constraint. A feasible solution is one that satisfies all constraints, and an infeasible solution is one that violates a constraint.

\subsection{The Quadratic Assignment Problem in E-commerce Websites}
\label{PREVIOUS STUDIES-The Quadratic Assignment Problem in E-commerce Websites}
Nishimura et al. \cite{nishimura2019item} proposed the following QAP for optimizing the order of items on an E-commerce website, in consideration of their diversity:
\begin{equation}
\label{eq:QAP_original}
\begin{array}{ll}
\text { maximize } & \sum_{i \in \text{Item}} \sum_{j \in \text{Site}} s_{i j} q_{i j} -w \\& \sum_{i \in \text{Item}} \sum_{i^{\prime} \in \text{Item}} \\& \sum_{j \in \text{Site}} \sum_{j^{\prime} \in \text{Site}} f_{i i^{\prime}} d_{j j^{\prime}} q_{i j} q_{i^{\prime} j^{\prime}} \\
\text { subject to } & \sum_{i \in \text{Item}} q_{i j}=1, \quad \forall j \in \text{Site},\\
& \sum_{j \in \text{Site}} q_{i j}=1, \quad \forall i \in \text{Item}, \\
& q_{i j} \in \lbrace0,1\rbrace, \quad \forall i \in \text{Item}, \quad \forall j \in \text{Site}.
\end{array} 
\end{equation}
Here the first term of the objective function represents estimated sales without considering similarity, the second is a penalty term to avoid sequences of highly similar items, “Item” is the set of items handled by the website, and “Site” is a set of positions where items are placed on the website. $q_{ij}$ is a binary variable that takes a value of 1 when item $i$ is placed in position $j$, $s_{ij}$ is estimated sales when item $i$ is in position $j$, $f_{ii^{\prime}}$ is similarity between items $i$ and $i^{\prime}$, $d_{jj^{\prime}}$ takes a value of 1 if positions $j$ and $j^{\prime}$ are adjacent and 0 otherwise, and $w$ is a penalty parameter.

In other words, the first term of the objective function in (\ref{eq:QAP_original}) represents estimated sales, and the second term penalizes orders of highly similar items. The first constraint is that each item must be placed in one position, and the second is that an item must occupy each position. By solving the optimization problem in (\ref{eq:QAP_original}), it is possible to find an item arrangement that maximizes sales while avoiding sequences of similar items. Formulating (\ref{eq:QAP_original}) as QUBO, we obtain
\begin{equation}
\label{eq:QAP_qubo}
\begin{array}{ll}
\text{minimize} & -\sum_{i \in \text{Item}} \sum_{j \in \text{Site}} s_{i j} q_{i j} \\&+w \sum_{i \in \text{Item}} \sum_{i^{\prime} \in \text{Item}}\\& \sum_{j \in \text{Site}} \sum_{j^{\prime} \in \text{Site}} f_{i i^{\prime}} d_{j j^{\prime}} q_{i j} q_{i^{\prime} j^{\prime}} \\
& +\lambda \sum_{i \in \text{Item}}\left(\sum_{j \in \text{Site}} q_{i j}-1\right)^{2}\\& + \lambda \sum_{j \in \text{Site}}\left(\sum_{i \in \text{Item}} q_{i j}-1\right)^{2} \\
\text { subject to } & q_{i j} \in \lbrace0,1\rbrace, \quad \forall i \in \text{Item}, \quad \forall j \in \text{Site},
\end{array}
\end{equation}
where $\lambda$ is a penalty parameter with a large value that makes the solution satisfy the constraints. We can optimize (\ref{eq:QAP_qubo}) by quantum annealing, but as explained in Section \ref{PREVIOUS STUDIES-Quantum Annealing Using D-Wave2000}, the one-hot constraints make couplers between logical variables fully connected, reducing the maximum size of solvable problems.
\subsection{Breaking One-hot Constraints}
\label{PREVIOUS STUDIES-Breaking One-hot Constraints}
In the QUBO formulation, we express constraints as a penalty term in the objective function. Especially in problems with multiple one-hot constraints, such as QAP and TSP, penalty terms create connections between arbitrary logical variables. As described in Section \ref{PREVIOUS STUDIES-Quantum Annealing Using D-Wave2000}, the graph structure of a quantum annealer has sparse connections between qubits, so it is necessary to use multiple qubits to represent one logical variable. As a result, D-Wave Advantage can only solve QAP up to size 12. This is a significant obstacle to optimization using quantum annealing. Ohzeki \cite{masayuki2020breaking} proposed a method, which we call the Ohzeki method in this paper, for efficiently solving combinatorial optimization problems with linear equation constraints such as one-hot constraints using a quantum annealer. The Ohzeki method converts the quadratic term generated by the penalty term to a linear term. As a result, the connections between qubits become sparse, making it possible to solve larger optimization problems with multiple one-hot constraints, such as QAP, with a quantum annealer, but conversely making it difficult to obtain a feasible solution. Briefly,
\begin{equation}
\label{eq:HS_QAP}
\begin{array}{ll}
\text { minimize } & f_{0}(\mathbf{q})\\
\text { subject to } & F_{k}(\mathbf{q})=C_{k}, \quad \forall k \in \lbrace 1,2,\ldots,M \rbrace ,\\
& \mathbf{q} \in \lbrace0,1\rbrace^{L},
\end{array} 
\end{equation}
is a combinatorial optimization problem with equality constraints, where $\mathbf{q}=(q_1,q_2, \ldots,q_L)$ are binary variables, $L$ is the number of binary variables, $M$ is the number of constraints, $F_{k}(\mathbf{q})$ is a linear function of $\mathbf{q}$, and $f_0$ is the objective function. The constraint $F_{k}(\mathbf{q})=C_k$ is a linear equality constraint. We can regard this as one-hot constraint if we set $F_{k}(\mathbf{q})=\sum_{k=1}^{L}q_{k}$, $C_{k}=1$. For optimization by a quantum annealer, we transform (\ref{eq:HS_QAP}) into the QUBO problem
\begin{equation}
\label{eq:HS}
\begin{array}{ll}
\text{minimize} & f(\mathbf{q})=f_{0}(\mathbf{q})+\frac{1}{2}\sum_{k=1}^{M}\lambda \left(F_{k}(\mathbf{q})- C_{k} \right)^2 \\
\text { subject to } & \mathbf{q} \in\{0,1\}^{L},
\end{array} 
\end{equation}
where $\lambda$ is a penalty parameter, the second term represents a penalty to satisfy the constraints, and $F_{k}(\mathbf{q})=C_{k} \;\; \forall k$.

The solution from a quantum annealer follows a Boltzmann distribution. We applied the Hubbard–Stratonovich transformation \cite{hubbard1959calculation}\cite{stratonovich1957method} to the second term $\frac{1}{2}\sum_{k=1}^{M}\lambda \left(F_{k}(\mathbf{q})- C_{k} \right)^2$ in the objective function in (\ref{eq:HS}), thus obtaining
\begin{equation}
\label{eq:HS-fnew}
f_{\text{new}}(\mathbf{q};\mathbf{v})=f_{0}(\mathbf{q})- \sum_{k=1}^{M} v_{k}^{t} F_{k}(\mathbf{q}),
\end{equation}
where $v_k^t$ is a parameter and $t$ is an index representing the number of iterations of the algorithm. This eliminates the dense connections between logical variables caused by $\left(F_{k}(\mathbf{q})- C_{k} \right)^2$, resolving the obstacle of reducing the problem size caused by the sparse QPU graph structure (Figure \ref{fig:pegasus}).

The solution sampled by the quantum annealer for $f_{\text{new}}$ in (\ref{eq:HS-fnew}) follows the Boltzmann distribution
\begin{equation}
\label{eq:HS-Q}
Q(\mathbf{q})^{t}=\frac{1}{Z(\mathbf{v})}\exp{\left(-\beta f_{\text{new}}\left(\mathbf{q};\mathbf{v}^{t}\right)\right)},
\end{equation}
with partition function
\begin{equation}
\label{eq:HS-Z}
Z(\mathbf{v})^{t}=\sum_{\mathbf{q}} \exp{\left(-\beta f_{\text{new}}\left(\mathbf{q};\mathbf{v}^{t}\right) \right)},
\end{equation}
where $\beta$ is a reverse temperature parameter. Index $t$ in $Q(\mathbf{q})^{t}$, $Z(\mathbf{v})^{t}$, and $\mathbf{v}^{t}$ represents iteration $t$ of the algorithm. While this method eliminates the dense connections between logical variables caused by the linear equality constraints, doing so does not satisfy the linear equality constraint $F_{k}(\mathbf{q})=C_{k}$ of the original combinatorial optimization problem (\ref{eq:HS_QAP}). To satisfy the constraints, we adjust $\mathbf{v}$ as
\begin{equation}
\label{eq:HS-adusting_v}
v_{k}^{t+1} \leftarrow v_{k}^{t}+\eta\left(C_{k}-\left\langle F_{k}(\mathbf{q})\right\rangle_{Q^{t}}\right), \quad \forall k \in \lbrace 1,2,\ldots,M \rbrace,
\end{equation}
where $\left\langle F_{k}(\mathbf{q})\right\rangle_{Q^{t}}$ represents the expected value of $F_k$  under distribution $Q^{t}$, and $\eta$ is a positive parameter. Calculation of $\left\langle F_{k}(\mathbf{q})\right\rangle_{Q^{t}}$ requires the expected value and is not easily computed on a classical computer. However, we can use quantum annealers to efficiently perform this computation. Since solutions $(\mathbf{q}_1,\mathbf{q}_2,\ldots,\mathbf{q}_N)$ are sampled from the quantum annealer, $\left\langle F_{k}(\mathbf{q})\right\rangle_{Q^{t}}$ equals the average $\frac{1}{N}\sum_{i=1}^{N}F_{k}(\mathbf{q}_{i})$, where $N$ is the number of solutions sampled from the quantum annealer. We adjust $\mathbf{v}$ to produce feasible solutions according to (\ref{eq:HS-adusting_v}), but as the experiments described in Section \ref{RESULT AND DISCUSSION} show, all sampled solutions generated by the Ohzeki method are infeasible despite satisfying some one-hot constraints. We therefore propose a heuristic post-processing method in Section \ref{PROPOSED METHOD-Bit-flip Heuristic in Postprocessing}.
\section{PROPOSED METHOD}
\label{PROPOSED METHOD}
\subsection{Applying the Ohzeki Method to the QAP on an E-commerce Website}
\label{PROPOSED METHOD-Applying the Ohzeki Method to the QAP on an E-commerce Website}
We solve the optimization problem (\ref{eq:QAP_original}) on the E-commerce website described in Section \ref{PREVIOUS STUDIES-The Quadratic Assignment Problem in E-commerce Websites} with a size larger than that used in \cite{nishimura2019item}. However, as mentioned above, the optimization problem (\ref{eq:QAP_original}) is the QAP, and the one-hot constraint causes the problem size to shrink. Therefore, in \cite{nishimura2019item}, problems with up to only 8 items can be solved by the quantum annealer D-Wave 2000Q. Even if D-Wave Advantage is used in the same way, problems with 13 or more items cannot be solved. To address this problem, we apply the Ohzeki method to (\ref{eq:QAP_original}). As a result, while we did not obtain a feasible solution, we can embed a 20-item problem (\ref{eq:QAP_original}) into D-Wave Advantage. Further, if using our proposed method as described in Section \ref{PROPOSED METHOD-Bit-flip Heuristic in Postprocessing} with the Ohzeki method, we can obtain approximate solutions for problems with up to 19 items.

In this section, we describe how to apply the Ohzeki method to the optimization problem (\ref{eq:QAP_original}) on an E-commerce website. First, we transform (\ref{eq:QAP_qubo}), which is the QUBO representation of (\ref{eq:QAP_original}), into the notation of (\ref{eq:HS}). We respectively define $f_{0}$, $F_{i}^{\text{Item}}$, and $F_{j}^{\text{Site}}$ as (\ref{eq:f_0 in e-commerce}), (\ref{eq:large_f_item in e-commerce}), and (\ref{eq:large_f_item in e-commerce}). $F_{i}^{\text{Item}}$ represents constraints on items on the E-commerce website. Similarly, $F_{j}^{\text{Site}}$ represents constraints on the position on the website. $F_{i}^{\text{Item}}$ and $F_{j}^{\text{Site}}$ correspond to $F_k$ in (\ref{eq:HS}), and $\mathbf{q}$ is an array of binary variables. We define $n$ as the number of items and $n = |\text{Item}| = |\text{Site}|$ holds. Using
\begin{equation}
\label{eq:f_0 in e-commerce}
\begin{split}
    f_{0}(\mathbf{q})&=-\sum_{i \in \text{Item}} \sum_{j \in \text{Site}} s_{i j} q_{i j} \\& +w \sum_{i \in \text{Item}} \sum_{i^{\prime} \in \text{Item}} \sum_{j \in \text{Site}} \sum_{j^{\prime} \in \text{Site}} f_{i i^{\prime}} d_{j j^{\prime}} q_{i j} q_{i^{\prime} j^{\prime}},
\end{split}
\end{equation}

\begin{equation}
\label{eq:large_f_item in e-commerce}
F_{i}^{\text{Item}}(\mathbf{q})=\sum_{j \in \text{Site}}q_{ij}, \quad \forall i \in \text{Item},
\end{equation}

\begin{equation}
\label{eq:large_f_site in e-commerce}
F_{j}^{\text{Site}}(\mathbf{q})=\sum_{i \in \text{Item}}q_{ij}, \quad \forall j \in \text{Site},
\end{equation}
we can transform the QUBO representation (\ref{eq:QAP_qubo}) to the following, like (\ref{eq:HS}):
\begin{equation}
\label{eq:HS in e-commerce}
\begin{array}{ll}
\text{minimize} & f_{0}(\mathbf{q})+\frac{1}{2}\sum_{i}\lambda \left(F_{i}^{\text{Item}}(\mathbf{q})- 1 \right)^2 \\&+\frac{1}{2}\sum_{j}\lambda \left(F_{j}^{\text{Site}}(\mathbf{q})- 1 \right)^2\\
\text { subject to } & \mathbf{q} \in\{0,1\}^{n^2}.
\end{array} 
\end{equation}
For convenience, we define $(F_{1},\ldots,F_{2n})=( F_{1}^{\text{Item}},F_{2}^{\text{Item}},\ldots,F_{n}^{\text{Item}},F_{1}^{\text{Site}},F_{2}^{\text{Site}},\ldots,F_{n}^{\text{Site}})$ in (\ref{eq:HS in e-commerce}), obtaining
\begin{equation}
\label{eq:HS_better in e-commerce}
\begin{array}{ll}
\text{minimize} & f_{0}(\mathbf{q})+\frac{1}{2}\sum_{k=1}^{2n}\lambda \left(F_{k}(\mathbf{q})- 1 \right)^2\\
\text { subject to } & \mathbf{q} \in\{0,1\}^{n^2}.
\end{array} 
\end{equation}
Similar to the transformation from (\ref{eq:HS}) to (\ref{eq:HS-fnew}), we can obtain
\begin{equation}
\label{eq:f_new in e-commerce}
    f_{\text{newec}}(\mathbf{q};\mathbf{v})=f_{0}(\mathbf{q})-\sum_{k=1}^{2n}v_{k}F_{k}(\mathbf{q}),
\end{equation}
using the Hubbard–Stratonovich transformation \cite{hubbard1959calculation}\cite{stratonovich1957method}, which eliminates the quadratic term caused by the constraints.

As explained in Section \ref{PREVIOUS STUDIES-Breaking One-hot Constraints}, solutions from the quantum annealer follow a Boltzmann distribution. In particular, when $f_{\text{newec}}$ is an energy function, the Boltzmann distribution is
\begin{equation}
\label{eq:Q in e-commerce}
    Q(\mathbf{q})^{t}=Q\left(\mathbf{q} ; \mathbf{v}^{t}\right)=\frac{\exp \left(-\beta f_{\text{newec}}\left(\mathbf{q} ; \mathbf{v}^{t}\right)\right)}{Z(\mathbf{v}^{t})},
\end{equation}
and its partition function is
\begin{equation}
\label{eq:Z in e-commerce}
Z(\mathbf{v})^{t}=\sum_{\mathbf{q}} \exp{\left(-\beta f_{\text{newec}}\left(\mathbf{q};\mathbf{v}^{t}\right) \right)}.
\end{equation}
To avoid violating the constraints, we update $\mathbf{v}$ as
\begin{equation}
\label{eq:v in e-commerce}
    v_{k}^{t+1} \leftarrow v_{k}^{t}+\eta^t \left(1 -\left\langle F_{k}(\mathbf{q})\right\rangle_{Q^{t}} \right), \quad \forall k \in \lbrace 1,2,\ldots,2n \rbrace.
\end{equation}
To compute $\left\langle F_{k}(\mathbf{q})\right\rangle_{Q^{t}}$, we use the average $\frac{1}{N}\sum_{i=1}^{N}F_{k}(\mathbf{q}_{i})$ by applying $F$ to solutions $(\mathbf{q}_1,\mathbf{q}_2,\ldots,\mathbf{q}_N)$ sampled from the quantum annealer.

In the above, we resolved the difficulty of reducing the problem size by applying the Ohzeki method to optimization problem (\ref{eq:QAP_original}), which was proposed in \cite{nishimura2019item}. However, while solutions obtained by the Ohzeki method are all infeasible in our experiment in Section \ref{RESULT AND DISCUSSION}, they satisfy some of the one-hot constraints. Thus, we can efficiently transform infeasible solutions to feasible ones using the BFHA, which we propose in the next section.
\subsection{Bit-flip Heuristic in Postprocessing}
\label{PROPOSED METHOD-Bit-flip Heuristic in Postprocessing}
\begin{figure*}[!t]
\centering
\includegraphics[width=5.0in]{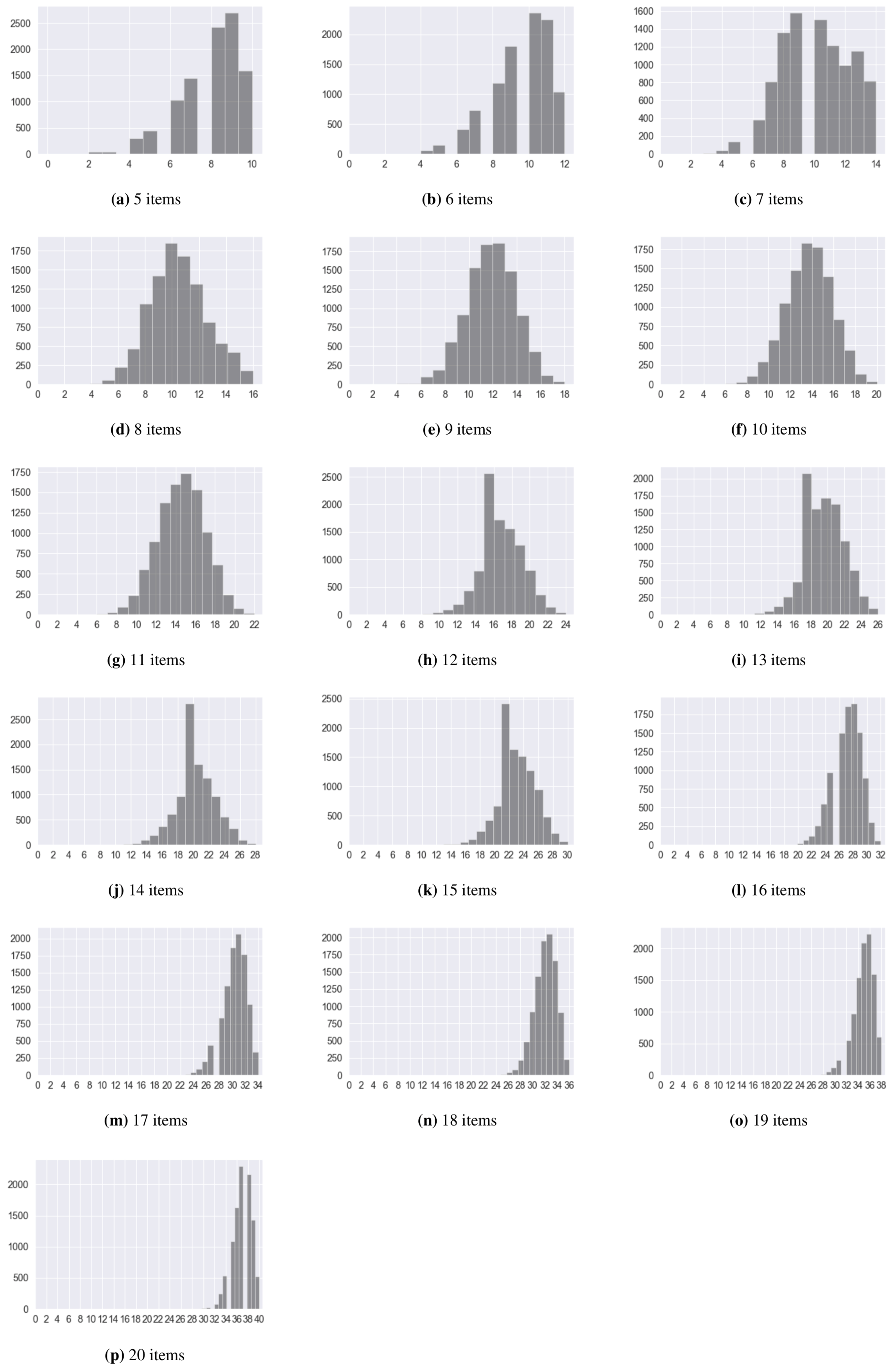}
\caption{The numbers of one-hot constraints violated in 10000 solutions obtained using the Ohzeki method with D-Wave Advantage. Vertical axes show the number of solutions, and horizontal axes show numbers of violated one-hot constraints. No problems had feasible solutions for all sizes.}             
\label{fig:ohzeki}
\end{figure*}

The Ohzeki method allows solving larger problems with one-hot constraints. However, when we optimize (\ref{eq:QAP_original}) using the Ohzeki method, we obtain no feasible solutions, as described in Section \ref{RESULT AND DISCUSSION}. Figure \ref{fig:ohzeki} shows the number of one-hot constraints violated in each of 10000 solutions generated by sampling 1000 solutions 10 times with D-Wave Advantage using the Ohzeki method. For example, a size-8 problem has 16 one-hot constraints. The number of violated one-hot constraints equals zero in a feasible solution. From Figure \ref{fig:ohzeki}, there are no solutions with 0 violations for any problem size, meaning all solutions are infeasible. Because it is difficult to obtain solutions satisfying all constraints using the Ohzeki method, we propose the BFHA, which transfers infeasible solutions violating one-hot constraints, such as (\ref{eq:QAP_original}), into nearby feasible solutions in terms of the Hamming distance. Before describing the BFHA, we show that the problem of transferring an infeasible solution to a nearby feasible solution is the assignment problem where $\Bar{q}_{ij}$ represents the infeasible solutions obtained by the Ohzeki method, and $q_{ij}$ represents the solution after moving by the minimum Hamming distance. The objective function consists of the sum of $1- q_{ij}$ for $(i,j)$ satisfying $\Bar{q}_{ij} =1$ and the sum of $q_{ij}$ for $(i,j)$ satisfying $\Bar{q}_{ij} =0$. In other words, the objective function represents the Hamming distance between $\Bar{q}_{ij}$ and $q_{ij}$. (\ref{eq:Hamming AP}) is thus the problem of finding a feasible solution $q_{ij}$ satisfying the minimum Hamming distance to $\Bar{q}_{ij}$. The Hungarian method \cite{kuhn1955hungarian}\cite{munkres1957algorithms} can similarly solve this problem as an assignment problem. By solving this optimization problem, an infeasible solution with constraint violations can be transferred to a feasible solution with the minimum Hamming distance. However, quantum annealing usually samples many solutions at once, so a faster algorithm is needed. We thus propose the BFHA, which is faster than these algorithms.

\begin{equation}
\label{eq:Hamming AP}
\begin{array}{ll}
\text { minimize } & \sum_{\Bar{q}_{ij}=1}(1-q_{ij}) + \sum_{\Bar{q}_{ij}=0}q_{ij}\\
\text { subject to } & \sum_{i \in \text{Item}} q_{i j}=1, \quad \forall j \in \text{Site},\\
& \sum_{j \in \text{Site}} q_{i j}=1, \quad \forall i \in \text{Item}, \\
& q_{i j} \in\{0,1\}, \quad \forall i \in \text{Item}, \quad \forall j \in \text{Site},
\end{array} 
\end{equation}

The following briefly describes the BFHA. In the case of problems with multiple one-hot constraints, such as the QAP and TSP, the solutions can be arranged as a square matrix. The sum of each row and column is $1$ in a feasible solution. Also, a solution is feasible if and only if $\sum_{i=1}^{n}q_{il}+\sum_{j=1}^{n}q_{kj} -2=0 ,\; \forall k,l$ holds. $\sum_{i=1}^{n}q_{il}+\sum_{j=1}^{n}q_{kj} -2$ represents difference from the feasible solution, which is expressed in line 4 of the BFHA (Algorithm 1). When the input solution to the BFHA is infeasible, the algorithm attempts to eliminate this gap.

Focusing on lines 6–9 of the algorithm, if $Violation$ has elements with a value of $1$ or more, the algorithm chooses the $(i,j)$ with the largest $Violation_{ij}$ satisfying $q_{ij}=1$. Then, the algorithm transforms $q_{ij}=1$ to $q_{ij}=0$ at $(i,j)$ and updates $Violation$. If no element of $Violation$ has a value of $1$ or more, and there are elements whose value is less than or equal to $-1$, then the algorithm executes lines 10-13. The algorithm chooses the $(i,j)$ with the smallest $Violation_{ij}$ satisfying $q_{ij}=0$ and transforms $q_{ij}=0$ to $q_{ij}=1$ at $(i,j)$. By repeating this procedure, an infeasible solution can be shifted to a nearby feasible solution in terms of the Hamming distance.

\begin{figure}[!t]
\centering
\includegraphics[width=3.3in]{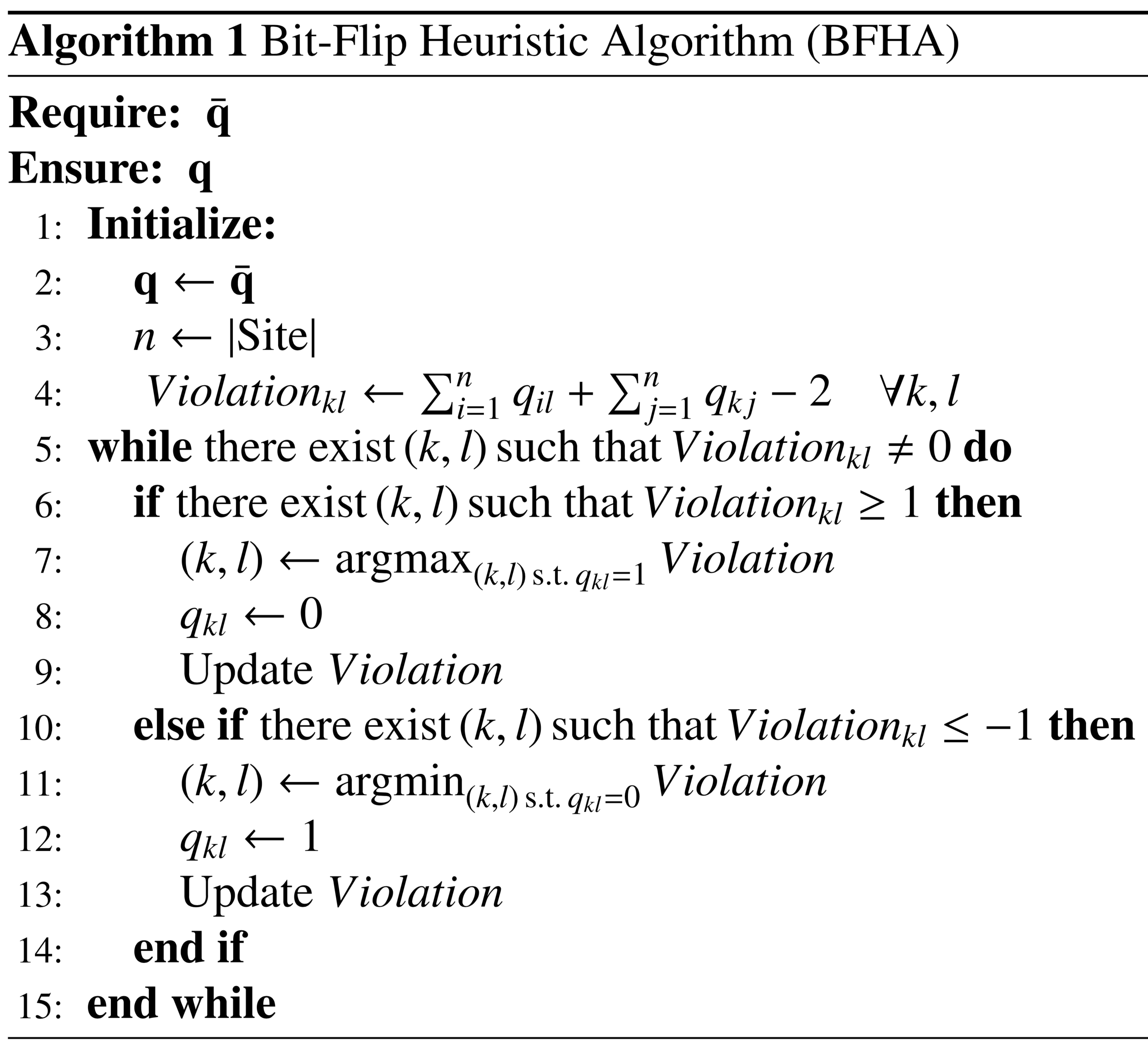}
\caption{Bit-Flip Heuristic Algorithm (BFHA)}             
\label{alg1}
\end{figure}

\begin{figure}[!t]
\centering
\includegraphics[width=3.3in]{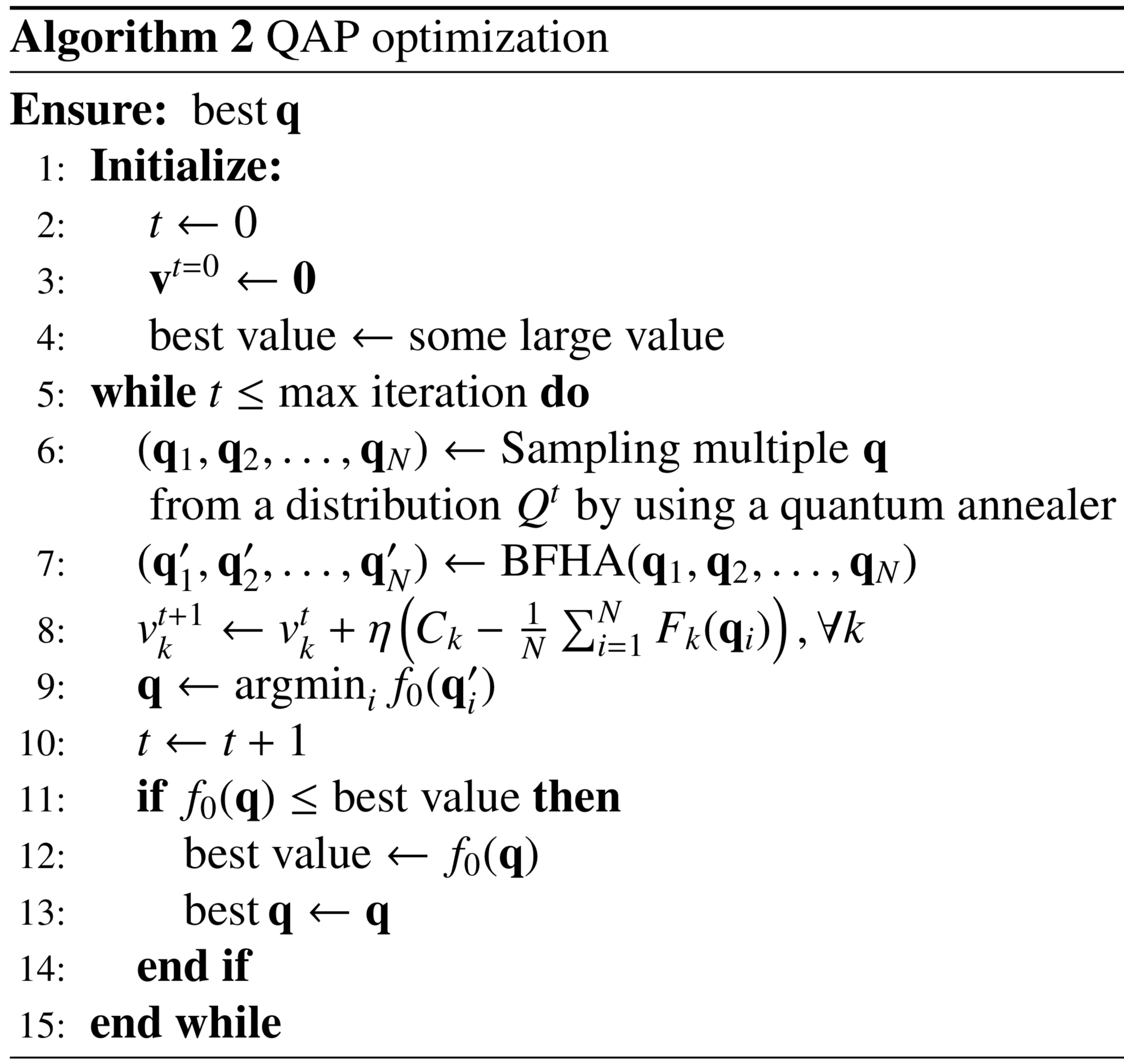}
\caption{QAP optimization}             
\label{alg2}
\end{figure}
To show that the BFHA can always map infeasible solutions to feasible ones, we first define $X := [\mathbf{q}_{ij}]$, an $n \times n$ matrix, $V := [\mathbf{v}_{ij}]$ where $\mathbf{v}_{ij} =\sum_{l=1}^{n}q_{il}+\sum_{k=1}^{n}q_{kj}-2$, $U := X \circ V $ with $\circ$ denoting the Hadamard product, and $D := (\mathbf{1}- X) \circ V$ where $\mathbf{1}=[1]_{n\times n}$.
\\
\begin{theorem}
The BFHA can always map infeasible solutions to feasible solutions.
\end{theorem}

\begin{IEEEproof}
$\mathbf{q}$ is a feasible solution if and only if $V=\mathbf{0}$, and $V=\mathbf{0}$ if and only if $(U=\mathbf{0}) \land (D=\mathbf{0})$. Next, we show that $\sum_{ij}U_{ij}$  is strictly reduced at lines 6-9 of the algorithm. We write $U$ as $U^t$ to clearly represent the $t$-th loop. If the largest element of $U^{t}$ is $(i^{\prime}, j^{\prime})$, lines 6-9 of the algorithm are equivalent to setting $U_{i^{\prime} j^{\prime}}^{t}=0$. Thus, using the fact that $U^{t+1}_{i^{\prime} j^{\prime}} =0 \times v_{i^{\prime} j^{\prime}}^{t+1}=0$, we have $U^{t}_{i^{\prime} j^{\prime}}-U^{t+1}_{i^{\prime} j^{\prime}}=U^{t}_{i^{\prime} j^{\prime}} \geq 1 $. From the above, lines 6-8 of the BFHA strictly reduce $\sum_{ij}U_{ij}$.

We next show that lines 10-13 of this algorithm do not generate elements of $U$ that are $1$ or more. In other words, we prove that lines 10-13 do not violate $U=0$. When the smallest element in $D$ satisfies $D_{ij}=-2$, no element equals $1$ in the $i$-th row and $j$-th column of $X$. Therefore, the operation at line 12 on $(i,j)$ satisfying $D_{ij}=-2$ has no effect on $U$.We show that the minimum value of the element of $D$ does not become $-1$ when $U=0$. We prove this by contradiction. Suppose the value of the smallest element of $D$ is $-1$. Then for the minimum element $(i,j)$ of $D$, $1$ always exists in the row or column containing $(i,j)$ of $X$. That is, there are $n$ or more elements in $X$ that equal $1$, but $U=0$ only when the algorithm is finished, which is a contradiction. From the above, lines 10-13 of this algorithm do not violate $U=0$. This algorithm strictly reduces $\sum_{ij}U_{ij}$ by lines 6-9 of the algorithm. Thus, $U=0$ will always be true. For the same reason, $D=0$ is always satisfied by lines 10-13. Also, $U=0$ is not violated by lines 10-13 of this algorithm, thus establishing $(U=\mathbf{0}) \land (D=\mathbf{0})$. Therefore, the BFHA can always map infeasible solutions to feasible ones.
\end{IEEEproof}
As mentioned above, it is difficult to obtain feasible solutions using only the Ohzeki method. We thus propose Algorithm 2, which always yields feasible solutions by applying both the Ohzeki method and the BFHA. Specifically, parameter $v$ is updated according to the Ohzeki method, and the BFHA is applied to the obtained solution every time, preserving the solution with the lowest objective function value. We can thereby solve problems (\ref{eq:QAP_original}) with sizes 5 to 20 using D-Wave Advantage, as described in Section \ref{RESULT AND DISCUSSION}.

\section{RESULT AND DISCUSSION}
\label{RESULT AND DISCUSSION}
\subsection{BFHA Performance}
\label{RESULT AND DISCUSSION-BFHA Performance}
We conducted experiments to demonstrate the BFHA speed and its ability to transfer infeasible solutions to nearby feasible solutions. The computing environment was an Intel Core i7-7700HQ 2.80 GHz CPU with 4 cores and 8 threads. Using the Ohzeki method, we generate 1000 infeasible solutions for each size of problem (\ref{eq:QAP_original}). We measured the following two indicators:
\begin{itemize}[\IEEEsetlabelwidth{X}]
\item Computation time (in seconds) per 1000 infeasible solutions
\item Hamming distance between an infeasible solution and a transferred feasible solution
\end{itemize}
For the above two indices, we conducted 10 experiments using the following three methods:
\begin{itemize}[\IEEEsetlabelwidth{X}]
\item Solve (\ref{eq:Hamming AP}) using Gurobi \cite{gurobi}
\item Solve (\ref{eq:Hamming AP}) by the Hungarian method \cite{kuhn1955hungarian}\cite{munkres1957algorithms}, using the Munkres implementation for Python \cite{murkres}
\item Use the BFHA
\end{itemize}

Table \ref{tab:heuristic} summarizes mean computation times and errors in Hamming distance, with ``Gurobi time [\si{\second}],'' ``Hungarian time [\si{\second}],'' and ``Bitflip time [\si{\second}]'' respectively showing computation times in seconds under the Gurobi, Hungarian, and BFHA methods. The ``Gap [\si{\percent}]'' column shows error between the minimum Hamming distance and that under the BFHA, indicating high accuracy. This table also confirms that the BFHA is much faster than the other algorithms, and that the increase in computation time for each size is larger with the Hungarian method than with the BFHA. However, the Gurobi computation time appears to increase at a slower rate, because it appears to take a long time to formulate the problem, thus gradually increasing calculation times. After building Violation, the BFHA performs only simple calculations such as adding and subtracting elements in matrix rows and columns. The BFHA can reduce the number of iterations and can find a feasible solution more quickly when there are few one-hot constraints violated among the infeasible solutions. Since some one-hot constraints are satisfied using the Ohzeki method, the BFHA is considered to be fast at finding a feasible solution.
\begin{table*}[!t]
\renewcommand{\arraystretch}{1.3}
\caption{Comparison of speed and Hamming distance}
\label{table_example}
\centering
\begin{tabular}{c c c c c}
\hline
 Size &  Gurobi time [\si{\second}]&  Hungarian time [\si{\second}]&  Bitflip time [\si{\second}]&  Gap [\si{\percent}]\\
\hline\hline
 5  & $3.991\times 10^{0}$ & $9.516\times 10^{-2}$ & $3.973\times 10^{-3}$ & $1.437$\\
 6  & $4.953\times 10^{0}$ & $1.605\times 10^{-1}$ & $4.476\times 10^{-3}$ & $2.256$\\
 7  & $5.560\times 10^{0}$ & $1.962\times 10^{-1}$ & $4.963\times 10^{-3}$ & $1.371$\\
 8  & $6.122\times 10^{0}$ & $2.236\times 10^{-1}$ & $5.886\times 10^{-3}$ & $1.712$\\
 9  & $6.743\times 10^{0}$ & $2.913\times 10^{-1}$ & $7.619\times 10^{-3}$ & $2.874$\\
 10 & $7.385\times 10^{0}$ & $3.393\times 10^{-1}$ & $7.836\times 10^{-3}$ & $2.588$\\
 11 & $8.096\times 10^{0}$ & $4.695\times 10^{-1}$ & $1.057\times 10^{-2}$ & $3.816$\\
 12 & $9.010\times 10^{0}$ & $5.171\times 10^{-1}$ & $1.167\times 10^{-2}$ & $4.159$\\
 13 & $9.784\times 10^{0}$ & $6.279\times 10^{-1}$ & $1.323\times 10^{-2}$ & $4.517$\\
 14 & $1.061\times 10^{1}$ & $6.930\times 10^{-1}$ & $1.404\times 10^{-2}$ & $4.596$\\
 15 & $1.173\times 10^{1}$ & $8.518\times 10^{-1}$ & $1.790\times 10^{-2}$ & $6.233$\\
 16 & $1.242\times 10^{1}$ & $1.061\times 10^{0}$  & $2.161\times 10^{-2}$ & $6.844$\\
 17 & $1.162\times 10^{1}$ & $1.117\times 10^{0}$  & $2.881\times 10^{-2}$ & $7.491$\\
 18 & $1.375\times 10^{1}$ & $1.151\times 10^{0}$  & $3.570\times 10^{-2}$ & $7.395$\\
 19 & $1.331\times 10^{1}$ & $1.237\times 10^{0}$  & $4.421\times 10^{-2}$ & $6.712$\\
 20 & $1.454\times 10^{1}$ & $1.434\times 10^{0}$  & $4.702\times 10^{-2}$ & $6.924$\\
\hline
\end{tabular}
\label{tab:heuristic} 
\vspace{-5mm}
\end{table*}

\subsection{Larger QAP Optimization on an E-commerce Website}
\label{RESULT AND DISCUSSION-Larger QAP Optimization on an E-commerce Website}
\subsubsection{Problem setting}
\label{RESULT AND DISCUSSION-Larger QAP Optimization on an E-commerce Website-Problem setting}
We randomly created 10 problems with 5 to 20 items and formulated them as (\ref{eq:QAP_original}). We set $w=0.5$ in (\ref{eq:QAP_original}), and set $\eta^{t}=0.1 \times 0.92^{t-1}$ in Algorithm 2 (except that $\eta^t = \eta^9$ if $t \geq 9$). We set the maximum number of iterations of Algorithm 2 to 30. We solved each problem one at a time using Algorithm 2. We used Advantage\_system1.1 with 5436 qubits, which is the latest version of D-Wave Advantage as of November 2020. We set quantum annealer parameters as num\_reads = 1000, annealing time = 50 \si{\micro \second}. We used the method of \cite{cai2014practical} to embed the problem in the quantum annealer.
\subsubsection{Result}
\label{RESULT AND DISCUSSION-Larger QAP Optimization on an E-commerce Website-Result}
Each problem was solved once by three methods: Algorithm 2, only the Ohzeki method, and a method that directly solves (\ref{eq:QAP_qubo}). Table \ref{tab:hresult} summarizes means of objective function values for the 10 best solutions by each method. Note that the Ohzeki method alone obtained no feasible solutions after 30 times with 1000 samples, so Table \ref{tab:hresult} does not list them. The ``With bitflip'' and ``QUBO'' columns respectively show the results of Algorithm 2 and the direct method. ``Score'' and ``Time [\si{\second}]'' are respectively means of the objective function value for the best obtained solution and the computation time, and ``Opt'' is the average of objective function values of the exact solutions. Cases where no feasible solution was obtained are denoted as ``infeasible,'' and problems of a size that cannot be embedded in the quantum annealer graph are denoted as ``cannot solve.''

We define the computation time for Algorithm 2 as the sum of the time for annealing, the BFHA execution time, the time for calculating expectations, and the time for updating parameters. Communication times between Japan and D-Wave Systems' quantum annealer in Canada is not included in the computation time. The computation time for ``QUBO'' includes only the time for quantum annealing. Calculations are terminated if an exact solution is found. Table \ref{tab:hresult} shows that Algorithm 2 (``With bitflip'') obtains overall good results. However, the difference between the obtained scores and ``Opt'' increases with size. For size 20 in particular, the solution accuracy deteriorated immediately. We found no feasible solutions by the Ohzeki method alone. We found no feasible solution for 6 out of 10 size-7 problems, so the average objective function value of the remaining four best solutions is given as the ``QUBO'' value. Similarly, we found no feasible solution for 8 out of 10 size-8 problems, and so use the remaining two solutions. We could obtain no feasible solutions for size 9 or more. From this, we can see that it is difficult to solve a problem with multiple one-hot constraints, such as the QAP, in the QUBO representation (\ref{eq:QAP_qubo}). In terms of speed, the computation times of Algorithm 2 show a gradual increase. The computation time of the quantum annealer does not change, and the increase in the BFHA computation time is not large. Therefore, the increase in calculation time is mostly due to line 8 in Algorithm 2.
\begin{table*}[ht] 
\caption{Results of a numerical experiment with D-Wave Advantage. } 
\centering 
\begin{tabular}{c cc  cc  c  } 
\hline 
 &\multicolumn{2}{c}{With bitflip} & \multicolumn{2}{c}{QUBO}\\
Size &  Score &      Time [\si{\second}] & Score  & Time [\si{\second}] &Opt\\ [0.25ex] 
\hline\hline
5  &     $-2.884\times 10^{0}$ & $1.165\times 10^{-1}$& $-2.870\times 10^{0}$  &  $7.200\times 10^{-1}$  &$-2.884\times 10^{0}$\\
 6  &     $-3.838\times 10^{0}$ & $2.856\times 10^{-1}$& $-3.588\times 10^{0}$  &  $1.385\times 10^{0}$   &$-3.838\times 10^{0}$\\
 7  &     $-4.779\times 10^{0}$ & $9.658\times 10^{-1}$& $-2.599\times 10^{0}$  &  $1.500\times 10^{0}$   &$-4.812\times 10^{0}$\\
 8  &     $-5.351\times 10^{0}$ & $2.901\times 10^{0}$ & $-1.944\times 10^{0}$  &  $1.500\times 10^{0}$   &$-5.530\times 10^{0}$\\
 9  &     $-5.949\times 10^{0}$ & $4.122\times 10^{0}$ & infeasible             &  infeasible             &$-6.245\times 10^{0}$\\
 10 &     $-6.308\times 10^{0}$ & $4.003\times 10^{0}$ & infeasible             &  infeasible             &$-6.906\times 10^{0}$\\
 11 &     $-6.964\times 10^{0}$ & $5.904\times 10^{0}$ & infeasible             &  infeasible             &$-7.880\times 10^{0}$\\
 12 &     $-7.307\times 10^{0}$ & $4.193\times 10^{0}$ & infeasible             &  infeasible             &$-8.612\times 10^{0}$\\
 13 &     $-7.876\times 10^{0}$ & $4.566\times 10^{0}$ & cannot solve           &  cannot solve           &$-9.629\times 10^{0}$\\
 14 &     $-8.231\times 10^{0}$ & $4.934\times 10^{0}$ & cannot solve           &  cannot solve           &$-1.040\times 10^{1}$\\
 15 &     $-8.778\times 10^{0}$ & $5.261\times 10^{0}$ & cannot solve           &  cannot solve           &$-1.133\times 10^{1}$\\
 16 &     $-9.169\times 10^{0}$ & $6.187\times 10^{0}$ & cannot solve           &  cannot solve           &$-1.211\times 10^{1}$\\
 17 &     $-9.482\times 10^{0}$ & $6.593\times 10^{0}$ & cannot solve           &  cannot solve           &$-1.307\times 10^{1}$\\
 18 &     $-9.915\times 10^{0}$ & $7.237\times 10^{0}$ & cannot solve           &  cannot solve           &$-1.156\times 10^{1}$\\
 19 &     $-1.042\times 10^{1}$ & $7.853\times 10^{0}$ & cannot solve           &  cannot solve           &$-1.209\times 10^{1}$\\
 20 &     $-1.078\times 10^{1}$ & $9.857\times 10^{0}$ & cannot solve           &  cannot solve           &$-1.563\times 10^{1}$\\  
\hline 
\end{tabular} 
\label{tab:hresult} 
\end{table*} 

\subsubsection{Discussion}
\label{RESULT AND DISCUSSION-Larger QAP Optimization on an E-commerce Website-Discussion}
The proposed method, Algorithm 2, showed good results overall. There are two possible reasons for this. First, by removing the one-hot constraint, connections between logical variables become sparse, so we can transform the original problem into one that is easy for the quantum annealer to solve. In \cite{hamerly2019experimental}, D-Wave 2000Q was used to solve the max-cut problem for regular graphs with different degrees. As the degree increases, that is, as the coupling becomes denser, the problem becomes more difficult to solve. Because $d$ in (\ref{eq:QAP_original}) makes problems sparse and we remove the one-hot constraints, we can transform the original problem into an easier problem for the quantum annealer. Secondly, the BFHA transfers infeasible solutions produced by the quantum annealer to nearby feasible ones according to Hamming distance.

However, the accuracy of size-20 problems was worse, and the same trend is observed in size-21 and above problems. We assume there are three main reasons for this. First, the chain is long and the number of qubits used is increased. We find that as the problem size increases, the chain length and the number of used qubits increase in the Pegasus graph. A chain is qubits used to represent a logical variable. Table \ref{tab:hardwareadvantage} shows the average number of couplers and the average chain length for each problem size in the D-Wave Advantage. We can see that as the problem size increases, the chain length and number of required qubits increases. As reported in \cite{hamerly2019experimental}, solution accuracy is known to becomes worse as the number of network qubits increases, which may have affected the results. Secondly, quantum annealer noise may have prevented correct sampling. It is known that more quantum annealer noise worsens solution accuracy \cite{dwave2019noise}, and this occurs with D-Wave Systems’ quantum annealer.  The third possibility is that we may not be able to obtain a good solution with the Ohzeki method, which is a heuristic algorithm with no theoretical guarantees, so the algorithm may have deteriorated solution accuracy. Further analysis of this point is a subject for future study.
\subsection{Contribution}
Our study made two main contributions. First, we proposed a heuristic algorithm named the bit-flip heuristic algorithm (BFHA), which moved solutions violating the one-hot constraint to nearby feasible solutions in terms of the Hamming distance. We also showed that the algorithm always transferred infeasible solutions to feasible ones. Furthermore, we numerically confirmed that the algorithm was fast and had high accuracy, meaning a small Hamming distance between the feasible solution assigned by the algorithm and the original infeasible solution. Second, we proposed a method for solving the QAP with quantum annealing by applying the BFHA to solutions obtained by the Ohzeki method. In a numerical experiment, we applied the proposed method to solve a sparse QAP that was used in areas such as item listing on an E-commerce website \cite{nishimura2019item}. Nishimura et al. \cite{nishimura2019item} solved this problem with up to 8 items using D-Wave 2000Q. If D-Wave Advantage had been used, problems with up to 12 items could have been solved. In this study, we successfully solved a QAP of size 19 with high accuracy for the first time using a quantum annealer without splitting.
\begin{table*}[ht]
\caption{Numbers of logical variables, required qubits, and required couplers, and average chain length for each problem size in the D-Wave Advantage.}
\centering
\begin{tabular}{lrrrr}
\hline
Size &  Logical variables &  Required Qubits &  Required Couplers &  Average chain length \\
\hline\hline
 5  &                 25 &               32 &                 80 &      1.280 \\
 6  &                 36 &               53 &                150 &      1.472 \\
 7  &                 49 &               83 &                252 &      1.694 \\
 8  &                 64 &              123 &                392 &      1.922 \\
 9  &                 81 &              176 &                576 &      2.173 \\
 10 &                100 &              248 &                810 &      2.480 \\
 11 &                121 &              325 &               1100 &      2.686 \\
 12 &                144 &              412 &               1452 &      2.861 \\
 13 &                169 &              529 &               1872 &      3.130 \\
 14 &                196 &              634 &               2366 &      3.235 \\
 15 &                225 &              784 &               2940 &      3.484 \\
 16 &                256 &              958 &               3600 &      3.742 \\
 17 &                289 &             1223 &               4352 &      4.232 \\
 18 &                324 &             1472 &               5202 &      4.543 \\
 19 &                361 &             1710 &               6156 &      4.737 \\
 20 &                400 &             2001 &               7220 &      5.003 \\
\hline
\end{tabular}
\label{tab:hardwareadvantage}
\end{table*}

\section{CONCULUSION}
In this study, we obtained approximate solutions for a sparse QAP of size 19 using D-Wave Advantage. We also proposed the BFHA, a method for moving infeasible solutions to a nearby feasible solution in terms of the Hamming distance. We also proved that the algorithm always moves infeasible solutions to feasible solutions, and numerically confirmed that the algorithm can move to the nearest feasible solution. We confirmed that the algorithm is faster than the compared algorithms. In a practical use case, we optimized item placement on an E-commerce website because this sparse QAP has been used in areas such as item listing on an E-commerce website. In \cite{nishimura2019item}, D-Wave 2000Q solved problems of up to size 8. When using the same method with D-Wave Advantage, it is impossible to solve problems of size 13 or larger. In addition, the Ohzeki method proposed in \cite{masayuki2020breaking} provided no feasible solutions to our problem. In contrast, using the BFHA with the Ohzeki method allowed optimization of up to 19 items in D-Wave Advantage with good accuracy.

Our proposed method is applicable to the real world applications such as item listing on an E-commerce website\cite{nishimura2019item} and minimizing the backtrack of jobs in the Generalized Flowline (GFL)\cite{gong1999genetic}, where GFL is a flow line on which jobs flow downstream and the backtrack is the movement of the auto guided vehicle (AGV) against the ideal direction of GFL. Moreover,
many real-world optimization problems involve one-hot constraints, so the proposed method has a wide range of applications, not only to the QAP but also to other problems with one-hot constraints. In this sense, the proposed method can thus make significant contributions to the development of real-world applications using quantum annealers. We obtained approximate solutions to size-19 QAP by the proposed method, but it is known that such approximate solutions can be quickly obtained even on classical computers. Like other studies \cite{neukart2017traffic}\cite{nishimura2019item}, this method is therefore not practical at present. Even so, quantum annealer technology is rapidly advancing, from 128 qubits in 2011 to 5640 qubits in 2020. In addition, noise has been reduced and more accurate solutions can be found, and quantum annealer hardware will continue to evolve in the future. For this reason, it is possible that this method will be practical for real-world applications in the future. However, it is necessary in future studies to address the problem of worsening accuracy for problems of size 20 and above.

\newpage


%

\footnote[0]{All company or product names mentioned herein are trademarks or registered trademarks of their respective owners.}
\end{document}